# Narrowband and high-rate entangled photon-pair generation from a high-Q silicon microring resonator

ShoichiroYasui[1,2], Tomohiro Inaba[1], Hidetaka Nishi[3,4], Reina Kaji[2], Satoru Adachi[2], Xuejun Xu[1*], Haruki Sanada[1]

[1] NTT Basic Research Laboratories, Atsugi, Kanagawa 243–0198, Japan
[2] Graduate School of Engineering, Hokkaido University, Sapporo, Hokkaido 060–8628, Japan
[3] NTT Device Technology Laboratories, Atsugi, Kanagawa 243–0198, Japan
[4] NTT Nanophotonics Center, Atsugi, Kanagawa 243–0198, Japan

**ABSTRACT:** Entangled photon-pair sources are indispensable building blocks of quantum information-processing technologies. Among the available approaches, on-chip microresonators are particularly promising owing to their resonant enhancement, CMOS-compatible fabrication, and wafer-scale integration capabilities. In this study, we optimized the structure of silicon microring resonators to suppress sidewall scattering. As a result, we achieved an intrinsic Q-factor of $1.26 \times 10^6$ using only standard fabrication processes. The high-Q resonator enabled a brightness coefficient of $3.9 \times 10^9$ Hz/GHz/mW$^2$, with a maximum brightness of 22.0 MHz/GHz and a maximum photon-pair generation rate of 9.19 MHz. Furthermore, Franson-type two-photon interference exhibited a visibility of $98.0 \pm 0.2\%$, confirming time–energy entanglement. These results show that narrow bandwidth and high generation rate can be achieved simultaneously in CMOS-compatible silicon photonic sources, advancing their use in quantum repeaters.

■ INTRODUCTION

Entangled photon-pair sources, generating photon pairs with strong quantum correlations, are essential for long-distance quantum information transfer [1]. In recent years, quantum information processing based on the quantum properties of entangled photons has progressed rapidly, and innovations are expected in diverse applications, including quantum communication [2,3], linear optical quantum computing [4,5], and quantum sensing and imaging [6–8].

Entangled photon pairs are typically generated via optical nonlinear processes such as spontaneous parametric down-conversion and spontaneous four-wave mixing (SFWM) [9]. As materials, both second-order nonlinear media (e.g., periodically poled lithium niobate [10,11] and aluminum nitride [12]) and third-order nonlinear media (e.g., silicon-on-insulator, SOI [13,14] and silicon nitride, $Si_3N_4$ [15]) have been extensively studied. Microresonators such as microrings (MRRs) and microdisk resonators (MDRs) enhance nonlinear interactions, enabling efficient photon-pair generation. As a result, compared with bulk or straight waveguide structures, they enable photon pairs to be generated at higher rates and with narrower spectral bandwidths.

High-rate and narrowband photon-pair sources are particularly important for quantum repeaters, which are central components of long-distance quantum communications and quantum networks. In quantum repeaters, entanglement swapping requires entangled photons to be stored in quantum memories (QMs) during the operation. Therefore, the spectral bandwidth of the photon pairs must be narrower than the inhomogeneous broadening of the optical transitions in QM materials [16,17]. For example, in the case of Er:$Y_2SiO_5$ crystals used for storing telecom-band photons, the inhomogeneous broadening is typically sub-GHz, depending on crystal quality [18,19]. Moreover, to compensate for system losses and the finite success probability of entanglement swapping, large photon-pair generation rates (PGRs) are required to enhance the scalability and throughput of the overall system [20–22]. Therefore, sources exhibiting high brightness—defined as the photon-pair generation rate per unit bandwidth—are highly desirable.

$Si_3N_4$ MRRs and AlGaAs MRRs are among the most successful integrated platforms for high-brightness photon-pair sources. For $Si_3N_4$ MRRs, the large bandgap makes two-photon absorption (TPA) negligible in the telecom band, enabling low-loss devices [23]. Resonance linewidths on the order of 10 MHz, with loaded Q factors of $5.0 \times 10^6$ and brightness coefficients of $1.2 \times 10^9$ Hz/GHz/mW$^2$, have been reported [24].

Furthermore, QM storage using narrowband photon pairs generated from $Si_3N_4$ MRRs has been demonstrated [25]. However, because the third-order nonlinearity of $Si_3N_4$ is smaller than that of silicon, achieving MHz-level PGRs typically requires ultra-high-Q, large-radius rings, which increase mode volume and impose stringent dispersion-engineering requirements for phase matching. For AlGaAs MRRs, the combination of a large third-order nonlinearity and low TPA has enabled brightnesses as high as $2 \times 10^{11}$ Hz/GHz/mW$^2$ using an AlGaAs-on-insulator MRR with a loaded Q factor of $1.24 \times 10^6$ [26]. Despite these advantages, AlGaAs still suffers from limited compatibility with standard CMOS processes. In contrast, silicon combines a strong third-order nonlinearity with full compatibility with standard CMOS fabrication. Taken together, these considerations motivate the optimization of Si-MRRs within standard CMOS-compatible foundry flows to realize photon-pair sources that simultaneously offer narrow bandwidths and high PGR.

Owing to the strong Kerr nonlinearity and the maturity of the silicon photonics platform, Si-MRRs have demonstrated MHz-level PGRs at pump powers as low as ∼10 μW, achieving record-high coincidence-to-accidental ratios (CARs) [14]. However, the reported photon-pair bandwidths have typically been on the gigahertz scale because the large Si/SiO$_2$ refractive index contrast leads to strong optical confinement and hence significant scattering due to sidewall roughness [27–29], and silicon at telecom wavelengths exhibits pronounced TPA and the resulting free-carrier absorption (FCA) at high pump powers [30], both of which make it difficult to realize ultra-high-Q Si-MRRs. Motivated by classical photonic applications such as narrowband filtering, microwave photonics, and on-chip lasers, high-Q Si-MRRs have therefore been extensively explored—for example, modified-Euler racetracks ($Q_{Load} = 1.6 \times 10^6$) [31] and adiabatic geometries combined with a local-oxidation (LOCOS) process ($Q_{Load} = 1.3 \times 10^6$) [32]—which achieve narrow resonances by suppressing roughness scattering. However, these geometries achieve high Q at the expense of a large mode volume V and ring perimeter, and thus are not expected to provide high PGRs. In the SFWM process, PGR is expressed as $R_{PG} \propto (Q_{Load}^3/V^2)\Phi_{PM}P^2$ [33], where $\Phi_{PM}$ is a phase-matching factor and $P$ is the pump power. In such designs, the increase in mode volume $V$ more than offsets the improvement in $Q_{Load}$, making it difficult to attain both narrow bandwidth and high PGR. Consequently, this trade-off between $Q_{Load}$ and the mode volume V remains a central obstacle to realizing high-brightness entangled photon-pair generation in Si-MRRs.

Here, we overcome the practical limitations imposed by the long-standing trade-off between $Q_{Load}$ and the mode volume V—and hence between narrow bandwidth and high PGR in Si-MRRs—and demonstrate that high-Q and high-brightness performance are simultaneously attainable in silicon without resorting to any nonstandard fabrication steps. By co-optimizing the microring radius and waveguide width, we minimize the field overlap with sidewalls and suppress scattering-induced losses, achieving an intrinsic Q-factor of $1.26 \times 10^6$ and a loaded Q-factor of $6.53 \times 10^5$ using standard foundry-compatible processes. This Q value represents one of the highest ever reported for Si-based entangled photon sources. As a result, our device achieves a narrow photon-pair bandwidth of ∼300 MHz, a brightness of 22.0 MHz/GHz, and a photon-pair generation rate of 9.19 MHz, along with Franson-type interference visibility of 98 ± 0.2%, confirming high-fidelity time–energy entanglement. This demonstration establishes a new benchmark for CMOS-compatible quantum photonics, showing that high-rate and narrowband entangled photon generation can be realized purely through intelligent device design rather than specialized post-processing. The approach not only advances silicon photonics as a practical quantum light-source platform but also provides a scalable path toward integrated quantum-repeater nodes capable of interfacing with solid-state quantum memories.

## ■ CHARACTERISTICS OF Si MICRORING RESONATORS

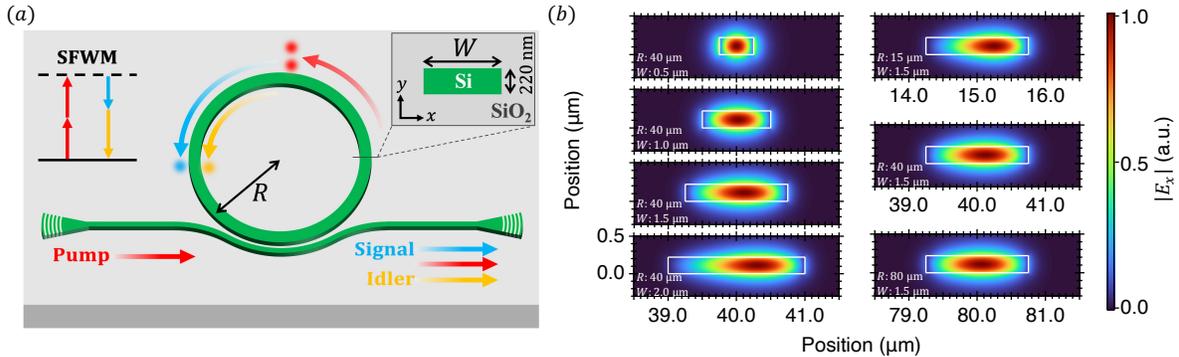

**Figure 1.** (a) Schematic diagram of a Si-MRR. The MRR consists of a ring of radius $R$ and width $W$, and a bus waveguide of width $W_{wg}$. The inset at the upper right schematically shows the ring cross section, where the transverse coordinates $(x, y)$ denote the radial (horizontal) and vertical directions, respectively, while the inset at the upper left illustrates the energy-conservation relation in the SFWM process. (b) Simulated electric field distribution of the fundamental TE mode in the ring cross section. The left column shows the case with fixed $R = 40$ μm and varying $W = 0.5$–$2.0$ μm, while the right column shows the case with fixed $W = 1.5$ μm and varying $R = 15$–$80$ μm.

To realize high PGR and brightness, we designed Si-MRRs with larger ring radii and widths compared with conventional structures, thereby suppressing roughness-scattering loss by shifting the electromagnetic field away from the sidewalls. Furthermore, we fabricated resonators with various values of $R$ and $W$ [Fig. 1(a)] and evaluated PGR and brightness to explore the optimal structure.

First, we simulated the electromagnetic field distribution of the fundamental TE mode ($TE_0$) in the cross section by using the finite-element method [Fig. 1(b)]. Reactive-ion etching (RIE) typically produces sidewall roughness on the order of a few nanometers [34]; therefore, the electric-field intensity in the ±5 nm sidewall surface region was evaluated. Figure 2(a) shows the dependence of sidewall field intensity on $W$ for $R = 15$, 40, and 80 μm. Because propagation loss due to roughness scattering arises from the interaction between the guided mode and the etched sidewalls and scales with the local field intensity at the rough interfaces [35], increasing $W$ and $R$ reduces the electric-field intensity near the sidewalls and thereby suppresses roughness-scattering loss. In particular, as $W$ increased from 0.5 to 1.5 μm, the sidewall field intensity at the inner ring side fell [Fig. 1(b)], and the overall sidewall field intensity decreased by more than one order of magnitude [Fig. 2(a)]. Moreover, the increase in $R$ from 15 μm to 80 μm further reduced the outer sidewall field intensity, resulting in an additional several-fold reduction in the total sidewall field intensity. These results suggest that resonators with larger radii and widths suppress the field near the sidewalls, thereby reducing roughness-scattering losses.

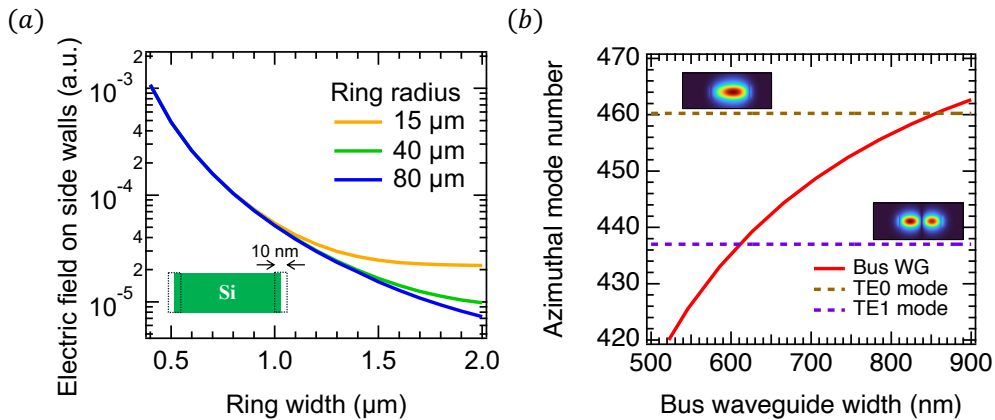

**Figure 2.** (a) Dependence of the normalized electric field intensity of the fundamental transverse electric ($TE_0$) mode near the inner and outer ring sidewalls (±5 nm from each sidewall) on the ring width $W$. Each color plot represents a different ring radius $R$. The inset shows the cross section of the Si-MRR. The black dotted line in the inset indicates the evaluation region. (b) Red curve shows the azimuthal mode number of the $TE_0$ mode as a function of bus waveguide width. Brown (purple) dashed lines denote the azimuthal mode number $m$ of the $TE_0$ ($TE_1$) mode for the MRR of $R = 40$ μm and $W = 1.5$ μm.

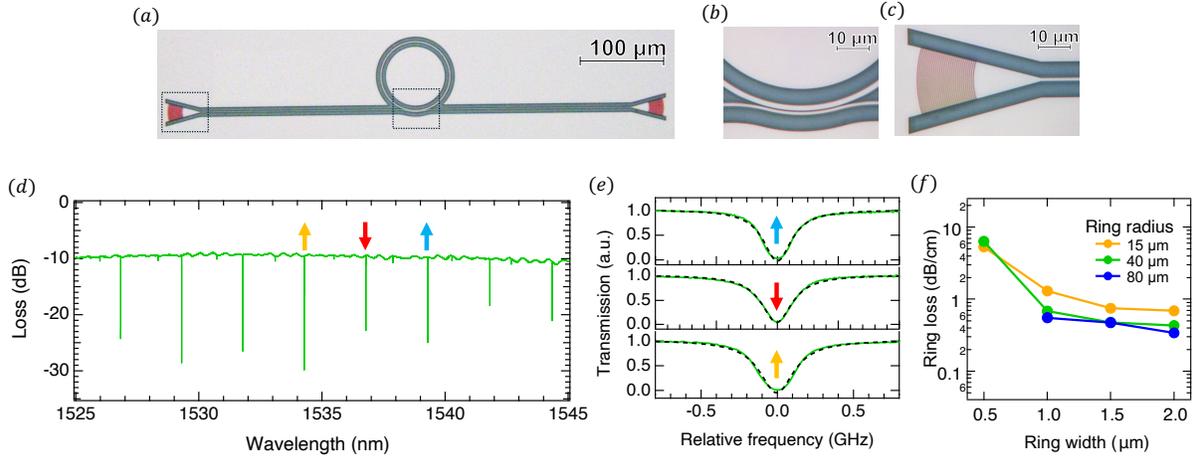

**Figure 3.** (a) Optical microscope image of a fabricated Si-MRR with $R = 40$ μm and $W = 1.5$ μm. (b), (c) Enlarged images of the directional coupler and grating coupler indicated by dashed boxes in (a). (d) Transmission spectrum of this MRR. Colored arrows denote the pump, signal, and idler resonances. (e) Enlarged views of these resonances. Black dotted lines indicate Lorentzian fits. (f) Dependence of ring propagation loss estimated from intrinsic Q on $R$ and $W$.

While increasing $W$ is beneficial for reducing sidewall-roughness scattering, when $W$ exceeds the single-mode regime, the waveguide becomes multimode. If the resonance frequency of a higher-order-mode lies close to that of the target $TE_0$ resonance, it can drive parasitic SFWM, generating unintended photon pairs and raising the noise floor. To eliminate this possibility, we optimized the bus waveguide width $W_{wg}$ to allow only $TE_0$ mode excitation via the directional coupler. In the coupler section, the ring and bus waveguide share coaxial arcs [Fig. 1(a)]. Their radii are $R$ and $R + W/2 + G + W_{wg}/2$, respectively, where $G$ is the gap between the ring and the bus waveguide. The azimuthal mode numbers are given by

$$m^{\text{ring}} = k_0 n_{\text{eff}}^{\text{ring}} R,$$

$$m^{\text{WG}} = k_0 n_{\text{eff}}^{\text{wg}} (R + W/2 + G + W_{wg}/2),$$

where $k_0$ is the wavenumber in vacuum and $n_{\text{eff}}$ is the effective refractive index of the ring or waveguide. Each propagation mode in the ring has its own $m$ ($m_{TE0}^{\text{ring}}$, $m_{TE1}^{\text{ring}}$, …) [Fig. 2(b)]. We designed $W_{wg}$ so that the $TE_0$ mode of the bus waveguide would be phase matched ($m^{\text{WG}} = m_{TE0}^{\text{ring}}$). As shown in Fig. 2(b), $W_{wg}$ was determined to be 856 nm for the ring with $R = 40$ μm and $W = 1.5$ μm. By optimizing the bus waveguide width, single-mode excitation of the $TE_0$ resonance was achieved.

Following the design described above, we fabricated Si-MRRs on an SOI wafer with a 220-nm-thick Si device layer using standard electron-beam lithography and RIE, and then clad them with $SiO_2$ deposited by chemical vapor deposition [Figs. 3(a) and 3(b)]. This straightforward process relies only on a single patterning and etching step followed by oxide deposition, without any additional sidewall-smoothing treatments such as thermal oxidation or resist reflow. Grating couplers [Fig. 3(c)] were used for optical input/output to fiber systems, with a coupling efficiency of 32% (5 dB) per coupler. Transmission spectra measured using a CW laser showed that higher-order modes were suppressed and that single $TE_0$ resonances close to critical coupling were predominantly observed [Fig. 3(d)]. Each resonance used for photon pair generation was fitted with a Lorentzian function,

$$T = \left| \frac{2i(f-f_0)/f_0 + 1/Q_{int} - 1/Q_{coupling}}{2i(f-f_0)/f_0 + 1/Q_{int} + 1/Q_{coupling}} \right|^2,$$

where $T$ is the transmittance, $f$ is the optical frequency, $f_0$ is the resonance frequency, and $Q_{int}$ and $Q_{coupling}$ are the intrinsic and coupling Q-factors, respectively. The loaded Q-factor is evaluated as

$$Q_{Load} = \left( Q_{int}^{-1} + Q_{coupling}^{-1} \right)^{-1} = \frac{f_0}{FWHM},$$

where FWHM is the full width at half maximum of the resonance. For the structure with $(R, W) = (40\ \mu m, 1.5\ \mu m)$, we obtained $Q_{int} = 1.26 \times 10^6$, $Q_{coupling} = 1.35 \times 10^6$, FWHM = 298 MHz(2.35 pm), and $Q_{Load} = 6.53 \times 10^5$. Furthermore, the propagation loss of the ring waveguide was estimated from the intrinsic Q-factor as

$$\alpha = \frac{2\pi f_0 n_g}{c Q_{int}},$$

where $c$ is the speed of light in vacuum and $n_g$ is the group index evaluated from the measured free-spectral range (FSR), yielding $\alpha = 0.526$ dB/cm. The transmission spectra and fitted parameters for MRRs with different $R$ and $W$ are summarized in Supporting Information A1. From these data, the propagation loss extracted from the intrinsic Q-factors [Fig. 3(f)] is found to decrease with increasing $R$ and $W$, in agreement with the trend of the simulated sidewall field intensity shown in Fig. 2(a). The intrinsic Q-factors obtained here are exceptionally high even among Si-MRRs (MDRs) reported to date [36,37]. These results show that mode linewidths (FWHM) on the order of a few hundred megahertz can be achieved in Si-MRRs fabricated by such a simple, foundry-compatible process, which is highly favorable for realizing bright, narrowband photon-pair sources.

### ■ PHOTON PAIR GENERATION MEASUREMENT RESULTS

Next, we evaluated the key performance metrics of the entangled photon-pair source—PGR and brightness by coincidence counting, and the visibility by Franson-type two-photon interference. A tunable CW laser in the telecom band was employed as the light source, and the frequency-modulation (FM) method [38] was used to lock the laser frequency to the resonance mode to stabilize the PGR [Fig. 4(a)] (see Supporting Information A2 for details). The laser was amplified by an EDFA and then passed through a narrowband (6 GHz) BPF to suppress broadband noise photons originating from spontaneous emission. The Si-MRR was temperature controlled and stabilized using a Peltier device. For the evaluation, CW pump light was coupled into the resonance mode at 1536.7 nm of a Si-MRR with $R$ = 40 μm and $W$ = 1.5 μm, generating photon pairs through the SFWM process. The generated signal and idler photons were separated from the pump by two cascaded WDM couplers with an extinction ratio of 80 dB. The residual pump light was monitored by an APD and used for FM feedback.

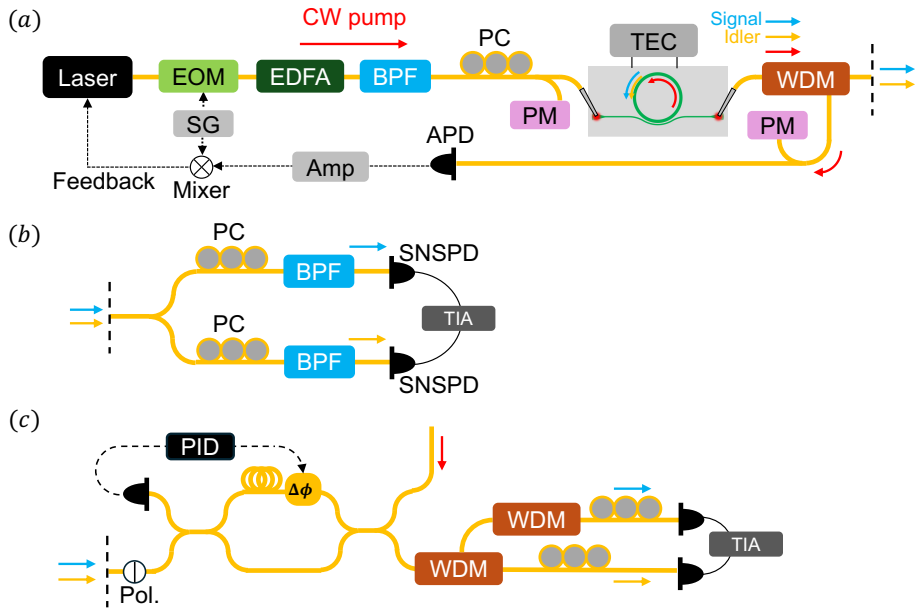

**Figure 4.** Schematic diagram of the experimental setup. (a) Experimental setup for photon-pair generation. (b) Coincidence measurement system between signal and idler photons. (c) Two-photon interference setup for measuring the visibility of time–energy entanglement. EDFA: erbium-doped fiber amplifier; EOM: electro-optic modulator; SG: signal generator; BPF: band-pass filter; PC: polarization controller; PM: power meter; TEC: thermoelectric controller; WDM: wavelength-division multiplexer; APD: avalanche photodiode; SNSPD: superconducting nanowire single-photon detector; TIA: time interval analyzer; Pol.: polarizer. These abbreviations denote the optical components used in the experiment.

In the coincidence counting, the signal and idler photons were divided by a 50:50 coupler and filtered by narrowband BPFs to select only the first-order photon-pair modes [Fig. 4(b)]. In the two-photon interference measurements, the photons were passed through linear polarizers and then injected into an unbalanced Mach–Zehnder interferometer (uMZI) [Fig. 4(c)]. A fiber stretcher was inserted into one arm to introduce a 15 m path-length difference, enabling variable relative phase. The relative phase was monitored by injecting a weak pump beam in the reverse direction and detecting single-photon interference with an APD, and a PID was used to lock it to an arbitrary value. The first-order photon-pair modes were subsequently separated by two different WDM couplers. In this experiment, the signal (idler) photons were detected by superconducting nanowire single-photon detectors (SNSPDs) with dark count rates of 316 (345) Hz, detection efficiencies of 59% (67%), and timing jitters of 58 (46) ps. A TIA with 42 ps jitter was used to record arrival-time correlations.

Figure 5(a) shows the single count rates of the signal and idler photons as a function of the on-chip pump power after the grating coupler, together with the coincidence count rates in a 4-ns window (right axis). The coincidence counts increased quadratically with pump power below 10 μW, confirming photon-pair generation from the SFWM process. At higher pump powers, both single and coincidence count rates exhibited saturation. To test whether the observed saturation arose from detector saturation, we inserted a 10-dB optical attenuator before the SNSPDs shown in Fig. 4(b); the same saturation trend persisted after accounting for the attenuation, thereby excluding SNSPD saturation as the cause. We therefore attribute the saturation to a device-internal mechanism— FCA generated by two-photon absorption TPA in silicon.

Next, time-correlation histograms were recorded with an integration time of 20 s and a bin width of 100 ps [Fig. 5(b)]. The vertical axis represents the second-order cross-correlation function defined as [9]:

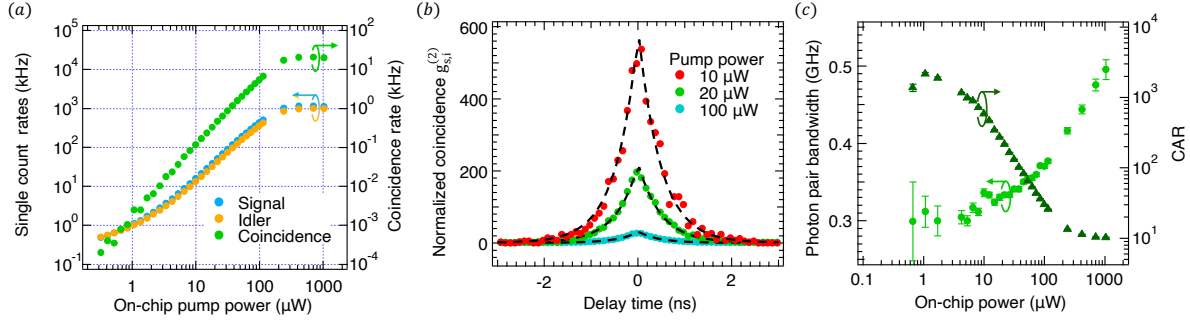

**Figure 5.** (a) Dependence of the single count rates of the signal and idler photons [left axis] and the coincidence count rate [right axis] on the on-chip pump power. (b) Time-correlation histograms at on-chip pump powers of 10 μW, 20 μW, and 100 μW. The vertical axis represents the second-order correlation function. (c) Dependence of the photon-pair bandwidth [left axis] and the CAR [right axis] on the on-chip pump power.

$$g^{(2)}_{s,i}(\Delta t) = \frac{R_{s,i}(\Delta t)}{R_s R_i \tau_w},$$

where $R_{s,i}(\Delta t)$ is the coincidence rate within a time window $\tau_w$ = 4 ns, and $R_s$, $R_i$ are the single count rates. The histograms at each pump power were fitted with

$$g^{(2)}_{s,i} = 1 + CAR e^{-|\Delta t|/\tau},$$

where $CAR$ denotes the coincidence-to-accidental ratio and $\tau$ is the cavity lifetime. $\tau$ is related to the photon-pair bandwidth $BW$ by $\tau = 1/(2\pi BW)$. The bandwidth estimated from $\tau$ in the low-pump regime was ~300 MHz [Fig. 5(c)], in good agreement with the FWHM obtained from the transmission spectrum. The increase in $BW$ with pump power confirms the reduction of Q-factor due to nonlinear absorption in Si. The maximum CAR at 1 μW pump was 2175 ± 100. There was a decrease in CAR due to SNSPD dark counts at low pump power and residual pump leakage and inelastic scattering at high pump power [39].

Next, we evaluated the on-chip PGR in the bus waveguide by correcting the coincidence counts for insertion losses from grating couplers and filters (Supporting Information A3). Figure 6(a) shows the experimental and calculated PGR for Si-MRRs with W = 1.5 µm and R = 15, 40, and 80 µm. The calculations accounted for the reduction in intrinsic Q due to nonlinear absorption in Si (Supporting Information A4). Figure 6(b) shows the pump-power dependence of brightness, calculated as PGR/BW using the BW dependence in Fig. 5(c), together with numerical results. Both PGR and brightness saturated with increasing pump power for all radii. The agreement between experiment and calculation indicates that the saturation originates from FCA induced by carriers generated by TPA. The slight deviations appearing in the high pump power regime are attributed to thermo-optic shifts [40] detuning the resonance from the pump frequency. In the low-power regime ($P < 10\ \mu W$), PGR and brightness coefficients were evaluated from quadratic fits, and the maximum experimental values are summarized in Table 1. For the structure with (R, W) = (40 µm, 1.5 µm), the PGR coefficient was $1286.7 \pm 21\ \text{MHz/mW}^2$ and the brightness coefficient was $3692 \pm 102\ \text{MHz/GHz/mW}^2$. The maximum measured values for this structure were PGR = 9.19 MHz and brightness = 22.0 MHz/GHz at $P = 240\ \mu W$. To the best of our knowledge, these values exceed previous records on CMOS-compatible platforms, including Si-MRRs [14] and Si-MDRs [37].

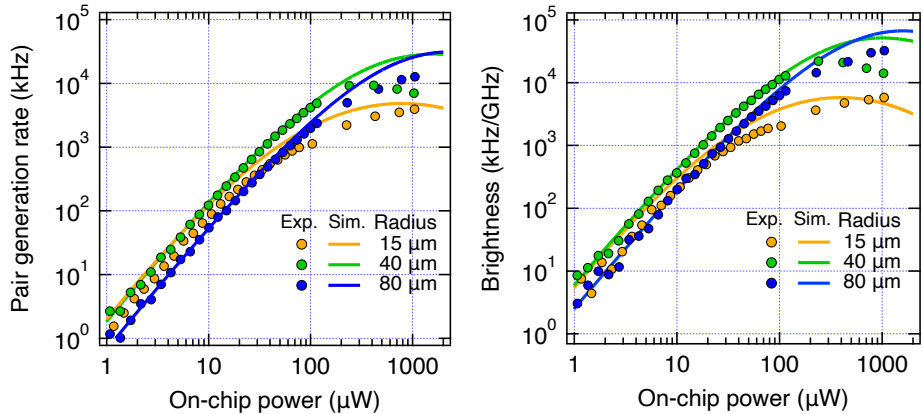

**Figure 6.** Pump-power dependence of (a) PGR and (b) brightness. Orange, green, and blue plots represent the measured results for resonators with $R$ = 15, 40, and 80 μm, respectively. The solid lines show the corresponding numerical calculations.

**Table 1.** Evaluated PGR and brightness coefficients, together with the maximum PGR and brightness values obtained.

| Ring radius (µm) | 15 | 40 | 80 |
|---|---|---|---|
| PGR coef. (MHz/mW$^2$) | 826 ± 19.7 | 1286.7 ± 21 | 539 ± 5.78 |
| Maximum PGR (MHz) | 3.9 | 9.19 | 12.7 |
| @ Power (µW) | @1046 | @240 | @1052 |
| Brightness coef. (MHz/GHz/mW$^2$) | 2050.5 ± 70.9 | 3962 ± 102 | 1946.6 ± 30.8 |
| Maximum Brightness (MHz/GHz) | 5.82 | 22.0 | 32.5 |
| @ Power (µW) | @1046 | @240 | @1052 |

While a narrow linewidth can, in principle, be obtained by increasing the Q-factors, the large PGR and brightness achieved here result from a favorable balance among the Q-factor, mode volume $V$, and phase-matching factor. As the ring radius R decreases, the scaling factors $Q^3/V^2$ (for PGR) and $Q^4/V^2$ (for brightness) increase, whereas the phase-matching factor decreases. Details are provided in Supporting Information A4. As a result of this trade-off, the maximum PGR and brightness were obtained at R = 40 µm rather than at R = 15 µm or 80 µm. These results indicate that maximizing performance in microring-based sources requires the simultaneous optimization of phase matching, mode volume, and the Q-factor.

Finally, for the resonator with (R, W) = (40 µm, 1.5 µm), which exhibited the maximum PGR and brightness coefficients, we verified time–energy entanglement by using Franson-type two-photon interference [41]. When the generated signal/idler pairs were injected into the uMZI [Fig. 4(c)], after the first coupler each photon was in a superposition of the short path $|S\rangle$ and the long path $|L\rangle$, and the overall state was

$$|\psi\rangle = \frac{1}{2}\left(|SS\rangle + e^{i\phi_i}|SL\rangle + e^{i\phi_s}|LS\rangle + e^{i(\phi_s+\phi_i)}|LL\rangle\right),$$

where $\phi_s$ and $\phi_i$ denote phase shifts of the signal/idler pairs in the long arms of the uMZI. At the second coupler, $|SL\rangle$ and $|LS\rangle$ yield side peaks separated by the interferometer delay $\Delta\tau$, while $|SS\rangle$ and $|LL\rangle$ overlap at zero delay. Postselecting the central peak gives the normalized two-photon state,

$$|\psi\rangle = \left(|SS\rangle + e^{i\Phi}|LL\rangle\right)/\sqrt{2}\,,\ \Phi = 2\Delta\phi + \phi_0,$$

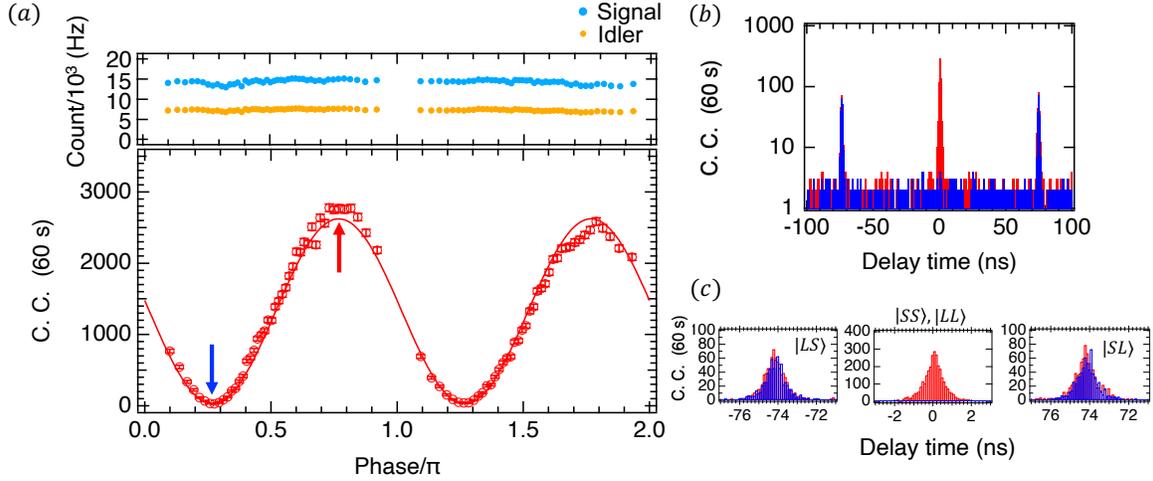

**Figure 7.** Results of the two-photon interference measurement at a pump power of 10 μW. (a) Relative-phase dependence of the single-count rates for the idler and signal photons (top) and of the coincidence count rate (bottom). The red and blue arrows indicate the minimum and maximum coincidence count points, respectively. (b) Time-correlation histograms measured at the constructive (red) and destructive (blue) interference phases. (c) Enlarged view of the time-correlation histogram showing the superposition of the $|SS\rangle$ and $|LL\rangle$ temporal modes, as well as the isolated $|LS\rangle$ and $|SL\rangle$ components.

where $\phi_0$ is the offset phase. Using projection bases $|\phi_{s,(i)}\rangle = (|S\rangle + e^{i\phi_{s,(i)}}|L\rangle)/\sqrt{2}$, the coincidence probability is

$$p = |(\langle\phi_s| \otimes \langle\phi_i|)|\psi\rangle|^2 = \{1 + \cos(2\Delta\phi + \phi_0)\}/4.$$

Thus, interference fringes at double the period appear as $\Delta\phi$ is scanned. The obtained fringes were fitted with a sinusoidal function, and visibility was evaluated as $V = (C_{\max} - C_{\min})/(C_{\max} + C_{\min})$, where $C_{\max\,(\min)}$ is the maximum (minimum) number of coincidence counts. The plot of the coincidence counts (2-ns window) versus $\Delta\phi$ in Figure 7(a) clearly shows that two-photon interference occurred while single counts remained phase independent, confirming suppression of first-order interference. At pump power of 10 μW, the raw visibility was $V = 0.980 \pm 0.002$. At pump powers of 50 μW and 100 μW, the visibilities were $0.949 \pm 0.006$ and $0.845 \pm 0.008$, respectively (see Supporting Information A5 for fringes). The time histograms at constructive/destructive phases [Figs. 7(b), (c)] show that the visibility reduction mainly originated from accidental coincidences. These results demonstrate that the Si-MRR source simultaneously achieves MHz-level PGR and ~10 MHz/GHz brightness while satisfying the Clauser-Horne-Shimony-Holt (CHSH) inequality violation condition [42], with $V > 1/\sqrt{2} = 0.707$, and that it functions as a narrowband entangled photon-pair source.

■ DISCUSSION

The Si-MRR achieved $Q_{\text{int}} = 1.26 \times 10^6$ and $Q_{\text{Load}} = 6.5 \times 10^5$ and was capable of narrowband photon-pair generation with a bandwidth of ~300 MHz under low-pump conditions. The PGR coefficient and brightness coefficient were $1.29 \times 10^9$ Hz/mW$^2$ and $3.96 \times 10^9$ Hz/GHz/mW$^2$, respectively. At a pump power of 240 μW, the maximum PGR and brightness reached 9.19 MHz and 22.0 MHz/GHz, respectively. This corresponds to several million entangled photon pairs per second within a ~300 MHz bandwidth at sub-milliwatt pump powers. Importantly, this bandwidth is narrower than the memory-bandwidth limit of atomic frequency comb (AFC) QMs based on $^{167}Er^{3+}$:$Y_2SiO_5$, where the usable AFC bandwidth is set by the smaller of the hyperfine level splittings (~1 GHz) and the inhomogeneous broadening of the optical transition , with reported values depending on crystal quality and including ~150 MHz [19], ~400 MHz [18], and in some cases, even exceeding the hyperfine splitting [25]. Consequently, the source can be directly bandwidth-matched to such memories without additional spectral filtering while still providing MHz-level entanglement-generation rates per spectral mode. Two-photon interference measurements confirmed raw visibilities exceeding $84.5 \pm 0.8\%$ for pump powers below 100 μW. Table 2 summarizes the key metrics of the photon-pair source at different pump powers. A comparison with previously reported MRR/MDR platforms shows that, particularly within CMOS-compatible Si-based platforms, our Si-MRR achieves state-of-the-art PGR and brightness over a wide range of pump powers (see Supporting Information A6). These results demonstrate that Si-MRRs fabricated in standard silicon photonics processes are capable of narrowband and high-rate operation and highlight their potential as entangled photon-pair sources for supporting high-throughput quantum-repeater links.

Table 2. Performance metrics of the Si-MRR photon-pair source with $(R, W) = (40\ \mu m, 1.5\ \mu m)$

| Pump power (μW) | $Q_{Load}$ | BW (MHz) | PGR (MHz) | Brightness (MHz/GHz) | CAR | Visibility (%) |
|---|---|---|---|---|---|---|
| 10 | $6.0 \times 10^5$ | 323 | 0.12 | 0.363 | 585 | $98.0 \pm 0.2$ |
| 50 | $5.5 \times 10^5$ | 354 | 1.85 | 5.21 | 70.5 | $94.9 \pm 0.6$ |
| 100 | $5.3 \times 10^5$ | 370 | 4.15 | 11.2 | 31.0 | $84.5 \pm 0.8$ |

On the other hand, the maximum CAR of $2175 \pm 100$, although high among reported Si-MRRs, did not reach the record $CAR_{max} = 12105 \pm 1821$ reported by Ma *et al.* [14]. This discrepancy may have originated from uncorrelated noise photons induced by inelastic scattering within the Si ring [39] or spontaneous Raman scattering in silica fibers/Si waveguides [37]. However, the latter cause is unlikely since spontaneous Raman scattering generally occurs in sidebands shifted by ~130 nm from the pump frequency [43]. To mitigate noise photons originating around the pump wavelength from inelastic scattering, higher-order pair modes can be exploited. In our resonator ($R = 40\ \mu m, W = 1.5\ \mu m$), the FSR was 2.4 nm, while Ma *et al.* used second-order pair modes with pump–idler/signal detuning of 19.8 nm. However, employing higher-order modes in high-Q MRRs narrows the phase-matching tolerance, potentially reducing the PGR. Thus, further optimization of the Si layer thickness and ring geometry for dispersion engineering (GVD control) will be required.

Although record-high PGR and brightness were demonstrated, further improvements are possible. Achieving ring structures with high Q, small mode volume V, and low GVD would further enhance these metrics. In particular, saturation induced by TPA/FCA can be mitigated by introducing a p-i-n junction around the ring and sweeping out carriers under reverse bias [44], thereby suppressing pump-dependent Q degradation and bandwidth broadening. Moreover, achieving lower GVD to improve phase matching is expected to further increase PGR and brightness in the low-pump regime [45]. In addition, improved fabrication precision would relax the Q-V trade-off, making the prospect of simultaneous optimization of the phase mismatch, mode volume, and Q-factor more realistic.

■ CONCLUSION

We demonstrated MHz-level PGR and ~10 MHz/GHz brightness while maintaining a narrow linewidth of ~300 MHz by using Si-MRRs fabricated entirely with standard processes. Furthermore, we confirmed high-visibility Franson-type two-photon interference under operating pump powers, evidencing time–energy entanglement. These results highlight the feasibility of implementing entangled photon-pair sources that simultaneously satisfy narrowband, high-rate, and high-fidelity requirements on CMOS-compatible Si photonics and pave the way toward quantum-repeater nodes integrating quantum memories for scalable quantum networks.

■ FUNDING

Japan Society for the Promotion of Science (JP23H01887, JP23K17883, JP23KJ0051, and JP25K21708).

■ ACKNOWLEDGMENTS

We thank Prof. Hideki Gotoh, and Prof. Junsaku Nitta for fruitful discussions.